\documentclass[prd,amsfonts,onecolumn,nofootinbib]{revtex4}

\usepackage{graphicx}

\newcommand{\be}{\begin{equation}} 
\newcommand{\ee}{\end{equation}}
\newcommand{\bea}{\begin{eqnarray}}
\newcommand{\eea}{\end{eqnarray}}

\newcommand{\gapp}{\mathrel{\raise.3ex\hbox{$>$}\mkern-14mu
              \lower0.6ex\hbox{$\sim$}}}
\newcommand{\lapp}{\mathrel{\raise.3ex\hbox{$<$}\mkern-14mu
              \lower0.6ex\hbox{$\sim$}}}

\newcommand\lsim{\lesssim}
\newcommand\gsim{\gtrsim}

\newcommand\vev[1]{{\langle {#1} \rangle}}
\renewcommand\({\left(}
\renewcommand\){\right)}
\renewcommand\[{\left[}
\renewcommand\]{\right]}

\newcommand\del{{\mbox {\boldmath $\nabla$}}}

\newcommand\eq[1]{Eq.~(\ref{#1})}
\newcommand\eqs[2]{Eqs.~(\ref{#1}) and (\ref{#2})}

\newcommand\eqreff[1]{(\ref{#1})}

\newcommand\pa{\partial}

\newcommand\mpl{M_{\rm P}}

\newcommand{\dlabel}[1]{\label{#1}}

\def\calp{{\cal P}}

\def\calpz{{\calp_\zeta}}

\newcommand\bfk{{\mathbf k}}

\newcommand\bfx{{\mathbf x}}

\newcommand\sub[1]{_{\rm #1}}

\newcommand\mone{^{-1}}
\newcommand\mtwo{^{-2}}
\newcommand\mthree{^{-3}}
\newcommand\mfour{^{-4}}
\newcommand\mfive{^{-5}}
\newcommand\mhalf{^{-1/2}}
\newcommand\half{^{1/2}}

\newcommand{\calpzchi}{\calp_{\zeta_\chi}}

\newcommand{\svev}{\sub{nl}}

\newcommand\modm[1]{|m(#1)|}
\newcommand{\mmt}{\modm{t}}
\newcommand{\mmts}{\modm{t}^2}
\newcommand{\kpeak}{k\sub{peak}}

\begin{document}

\title {Primordial black hole formation and hybrid inflation}

\author{David H. Lyth}

\affiliation{Department of Physics, Lancaster University, 
Lancaster LA1 4YB, UK}

\begin{abstract}
We calculate the contribution to the curvature perturbation, that is generated while
the waterfall field  grows exponentially
(which can occur only if the waterfall field mass is $m\gg H$).
We consider the upper bound on its spectrum coming from constraints
on the abundance of primordial black holes,  taking account
of possible non-gaussianity. The constraint is satisfied, but extrapolation
of our result to the regime 
$m\sim H$ suggests that the constraint might not be satisfied there.
\end{abstract}

%\keywords{Keywords}
%\arxivnumber{archive number}

\maketitle

\section{Introduction}

On cosmological scales, the primordial curvature perturbation
$\zeta$ is very small with spectrum $\calpz(k)\simeq (5\times 10\mfive)^2$.
But $\calpz(k)$ may rise as  the wavenumber $k$  increases, to some much bigger
peak value $\calpz(k\sub{peak})$.  If that happens,
primordial black holes may form as $k\sub{peak}$ enters the horizon, and 
cosmological bounds on their abundance translate to an upper bound
on $\calpz(k\sub{peak})$.

The bound is usually discussed under the assumption that $\zeta$ is nearly
gaussian, giving  very roughly $\calpz(k\sub{peak})\lsim 10\mtwo$. 
In the first part of this paper we discuss the bound assuming instead
that on scales around $k\sub{peak}$
\be
\zeta(\bfx) = \pm \( g^2(\bfx)- \vev{g^2} \)
, \dlabel{zetang} \ee
with $g$ gaussian. We show that the  bound becomes very roughly
$\calpz(k\sub{peak})\lsim 10\mthree$ 
for the positive sign and $\calpz(k\sub{peak})
\lsim 1$ for the negative sign.

It is known \cite{ourbh} that the form \eqreff{zetang} can hold  if 
the rise in $\calpz(k)$ is generated after inflation by a curvaton-type
 mechanism. For the curvaton mechanism itself the sign is positive 
but it could be negative more generally. With a  curvaton-type mechanism,
$\kpeak$ corresponds to the horizon scale when the mechanism ceases
to operate and black hole formation follows immediately.

A different possibility is for the rise to be generated by the 
 waterfall field of  hybrid inflation. 
A calculation has been done 
for a particular case \cite{p10}, where  simple expressions  were found
for $\zeta$ and $\calpz$. In that case $\zeta$ is gaussian and the black
hole bound is well satisfied.
The main purpose of this
 paper is to generalise the calculation of
\cite{p10}, to cover essentially any hybrid inflation model
permitting exponential growth of the waterfall field.

\section{Black hole bound}

\dlabel{sbh}

The bound that we are going to consider 
rests on the validity of the following statement: if,
 at any epoch after inflation,
there are  roughly spherical and horizon-sized regions with $\zeta$
significantly bigger than 1, a significant fraction of them  will collapse
to form roughly horizon-sized   black holes.\footnote
{As in \eq{zetang} we are choosing the background scale factor $a(t)$ so that
the perturbation $\zeta=\delta(\ln a(\bfx,t))$ has zero spatial average.}
The validity is suggested by 
 the following argument: the overdensity at horizon entry
is $\delta\rho/\rho \sim \zeta$, and if it is of order 1 then
 $\delta \rho \sim \rho=3\mpl^2H^2$. The excess energy within the Hubble distance
 $H\mone$
is then  $M\sim H\mthree \rho \sim \mpl^2/H$, which means
that the 
Hubble distance corresponds roughly to the Schwarzchild radius of a black
hole with mass $M$. The validity 
is confirmed by detailed calculation using several
different approaces, as summarised for instance in \cite{bhbound}.

Before continuing we mention the following caveat. Practically all of the literature,
as well as the simple argument just given, assumes
 that $\zeta$ within the region is not {\em very much} bigger than
1. Then the spatial geometry within the region is not too strongly distorted
and the size of the black hole is indeed roughly that of the horizon. 
In the opposite case, the background geometry is strongly distorted and the wavenumber
$k$ defined in the background no longer specifies the physical size of the region
at the epoch $aH=k$ of horizon entry   \cite{bhform2}. 
An entirely different discussion would then
be necessary, which has not been given in the literature. As the opposite case does
not arise in typical early-universe scenarios we ignore it.

We are interested in the case that $\calpz(k)$ has a peak at some value
$\kpeak$, and we assume that the width of the peak in $\ln k$ is roughly of order
1 so that $\vev{\zeta^2} \sim \calpz(\kpeak)$. 
We will focus on the case that 
 $\zeta$ is generated by the waterfall field perturbation. Then  
$\calpz(k)\propto k^3$
below the peak with an exponential  fall-off above the peak, giving\footnote
{If the integral fails to converge at small $k$ we impose a cutoff corresponding
to the size of the observable universe. The contribution from small $k$
then has a negligible effect.} 
\be
\vev{\zeta^2} \simeq  \int^\infty_0 \calpz(k) dk/k = \frac13 \calpz(\kpeak)
, \dlabel{kcubed} \ee
but the numerical factor is not very important.
What matters is that
 regions with $\zeta\gsim 1$ that might
form black holes will be rare if  $\calpz(\kpeak)$ is sufficiently far below 1, 

Observation demands that the regions must indeed be rare, because
it places a strong upper bound 
on the fraction of $\beta$ of space that can collapse 
to form black holes, on the assumption that the collapse takes place at a single
epoch as is the case in our scenario. 
A recent investigation of the bound  is given in 
\cite{bhbound}, with extensive references to the literature. The bound can arise from 
 many different physical effects, depending on the epoch of collapse, and is subject
to many uncertainties. Typical bounds  are roughly in the range
\be
10^{-20} < \beta\sub{max} < 10\mfive
, \dlabel{betabound} \ee
 and we shall take that to be the range in what follows.
To bound $\calpz(\kpeak)$, we 
 shall require $f<\beta\sub{max}$, where $f$ is the  fraction of space 
with  $\zeta>1$.

That fraction can be calculated from $\vev{\zeta^2}$
if we know the probability distribution of $\zeta(\bfx)$. The standard 
assumption is that it is gaussian. Then
\be
f = \mbox{erfc\,} (1/\sqrt 2 \vev{\zeta^2}\half )
, \ee
and using the large-$x$ approximation $\mbox{erfc\,}(x)\simeq e^{-x^2/2}$ we find
$\vev{\zeta^2} \simeq 1/2 \ln(1/f)$. 
For the range \eqreff{betabound} this gives (with \eq{kcubed})
 $\calpz(\kpeak)\lsim 0.01 $ to $0.04$.

If instead $\zeta$ 
has the non-gaussian form \eqreff{zetang} with the plus sign we have
\be
\vev{\zeta^2} =  2\vev{g^2}^2 \simeq  2 \[ \frac{1}{2 \ln(1/f) } \]^2
, \ee
which with \eq{kcubed} 
 gives $\calpz(\kpeak)\lsim 6\times 10\mfour$ to $5\times 10\mthree$.

The situation when $\zeta$ has the form \eqreff{zetang} with the minus
sign is quite different. There is   now  no region
of space where $\zeta > \vev{g^2}$, and 
 $f\ll 1$ now implies some bound $\vev{g^2}-1\ll 1$
which is practically equivalent to  $\vev{g^2}< 1$.
With \eq{kcubed} this  corresponds to $\calpz(\kpeak)< 6$.

\section{Evolution of the waterfall field}

\dlabel{sevolution}

We adopt the notation and basic approach of \cite{p10}. In the first two subsections
these are summarised, and then we present our new calculation of evolution of
the waterfall field $\chi$.

\subsection{Standard hybrid inflation}

We  consider  the  usual hybrid inflation  potential, 
\bea
V(\phi,\chi) &=& V_0 + V(\phi)
+\frac12 m^2(\phi)  \chi^2 + \frac14\lambda \chi^4
 \dlabel{fullpot}  \\
m^2(\phi(t)) &\equiv & g^2\phi^2(t)  -m^2 \equiv g^2 \(\phi^2(t) 
-\phi\sub c^2 \)
, \eea
with 
 $0< \lambda\ll 1$  and $0< g\ll 1$. We consider the standard case $m\gg H$. 
The effective mass-squared  $m^2(\phi(t))$ of the waterfall field goes negative
when the inflaton $\phi$ falls below $\phi\sub c\equiv m/g$. Then $\chi$ moves towards its vev,
marking the beginning of what is called the  waterfall. The waterfall is deemed to
end when $\chi$ approaches its vev, which usually marks the end of inflation.

At least after the observable universe leaves the horizon, the
 inflaton potential $V(\phi(t))$  is supposed to have positive slope, 
and the inflaton is supposed to have  zero vev so that $V(0)=0$.
Also, it is assumed that  $V(\phi)\ll  V_0$.

The requirements that $V$ and $\pa V/\pa \chi$ vanish in the vacuum
give the vev $\chi_0$ and the inflation scale $V_0\simeq 3\mpl^2 H^2$:
\be
\chi_0^2=\frac{m^2}\lambda\simeq 12\mpl^2 H^2/m^2,
\qquad V_0=\frac{m^4}{4\lambda }\simeq 3\mpl^2 H^2
 \dlabel{chiandv}. \ee

The waterfall is supposed to begin with an era during which the evolution of 
$\chi$ is linear.
 Choosing a gauge whose slicing corresponds to uniform $\phi$,
\be
\ddot \chi_\bfk + 3H \dot\chi_\bfk + \[ (k/a)^2 + m^2(\phi(t)) \] \chi_\bfk
= 0
. \dlabel{chiddot} \ee
The  energy density and pressure of  $\chi$ are
\bea
\rho_\chi &=&  m^2(\phi)\chi^2 + \frac12\dot\chi^2 + \frac12 |\del \chi|^2
\dlabel{rchi} \\
p_\chi &=& - m^2(\phi)\chi^2 + \frac12\dot\chi^2 + \frac16 |\del \chi|^2
. \dlabel{pchi} \eea

In  \cite{p10} we considered a regime of parameter space is which 
(i) the waterfall  takes much less than a Hubble time,  (ii)
$m^2(t)\propto t$  during the linear era  and 
 (iii) $\chi$ is growing exponentially
by the end of the linear era. We found simple formulas for 
 the contribution  $\zeta_\chi$ of $\chi$ to the curvature perturbation
to at the end of the linear era, 
and for its spectrum $\calpzchi$.
In this paper show that  very similar formulas hold if we assume only 
exponential growth.

\subsection{Exponential growth}

Since inflation by definition  continues during the 
the waterfall, $H$ will not vary much and we set it equal to a constant
to simplify the presentation.  Then 
 conformal time is $\eta=-1/aH$ and  \eqreff{chiddot} can  be written 
\be
\frac{d^2(a\chi_\bfk)}{d\eta^2} +\omega_k^2 a\chi_\bfk =0,  \dlabel{fulleq} 
\ee
with 
\be
\omega_k^2(\eta) \equiv  k^2 +a^2\tilde m^2(t),\qquad \tilde m^2\equiv  
m^2(t)-2H^2,\qquad m^2(t) \equiv g^2\phi^2(t) - m^2
 .  \ee

For sufficientlly small $k$, we can set $\omega_k^2\simeq  \omega_{k=0}^2=a^2\tilde m^2$.
Then $\omega_k^2$ switches  from positive to negative before  $\phi=\phi\sub c$, but presumably
not long before since $m\gg H$. For $k^2>0$ the switch is later.
For the scales that we need to consider, we   assume that 
 there  are eras  both before and after the switch 
when  $\omega_k^2$ satisfies the adiabaticity condition
$d|\omega_k^2|/d\eta  \ll |\omega_k^2|$. 

During the adiabaticity era before the switch we take the mode function to be
\be
a\chi_k \simeq  (2\omega_k(\eta))\mhalf \exp\( -i\int^\eta
 \omega_k(\eta) d\eta \)
, \dlabel{before} \ee
which defines the vacuum state.
During the adiabaticy era after the switch 
\be
a\chi_k \sim  (2|\omega_k(\eta)|)\mhalf \exp\( \int^\eta_{\eta_1(k)}
 |\omega_k(\eta)| d\eta \) \dlabel{after}
, \ee
where the subscript 1 denotes the beginning of the adiabatic era.
The displayed prefactor holds \cite{p10} only  if $m^2(t)\propto t$ and $H(t-t_1)\ll 1$
but its precise form doesn't matter. All we   need is for it to vary sufficiently
slowly that the growth is dominated by the exponential.
We call the era during which that is true the 
growth era.

During  the growth era,
 the adiabaticity condition 
is equivalent to the three  conditions
\bea
\tilde m(t) &\simeq&  m(t), \dlabel{adcon1}\\
\frac{2H}{\mmt}  &\ll&
\[ 1 - \(\frac k{a(t) \mmt } \)^2 \]^{3/2} \dlabel{adcon2} \\
 \frac1{\mmts}\frac{d\mmt}{dt} ,
 &\ll&
\[ 1 - \(\frac k{a(t) \mmt } \)^2 \]^{3/2}
. \dlabel{adcon3} \eea
The adiabatic era begins when all three conditions are first satisfied.

By virtue of \eqs{adcon1}{adcon2}, we have from \eq{after}  
$\dot \chi_k \simeq |m(t)| \chi_k$. Except near places where $\chi(\bfx,t)$
vanishes, this implies
\be
\dot \chi(\bfx,t) \simeq |m(t)| \chi(\bfx,t) . \dlabel{chidot}
. \ee

In the regime $k\ll a\mmt$ we have
\be |\omega_k| \simeq a\mmt \(1- \frac12 \frac{k^2}{a^2\mmts} \)
 \ee
giving 
\be \chi_k(t)  \simeq  \chi_{k=0}(t)  e^{-k^2/2k_*^2(t)},\qquad
 \chi_{k=0}(t)  \simeq (2a^3\mmt)\mhalf
\exp\( \int^t_{t_1} dt \mmt   \) 
, \ee
where 
\be
k_*^2(t) \equiv\(  \int^t_{t_1}\frac{dt}{a^2\mmt} \)\mone
.\ee

During the growth era, $\chi_\bfk(t)$ is classical and proportional
to the mode function $\chi_k(t)$. The same holds for $\chi(\bfx,t)$, except near
places where it vanishes. The spectrum is 
 $P_\chi= \chi_k^2$. Also,
\be
\vev{\chi^2(t} = \frac{4\pi}{(2\pi)^3}
 P_\chi(0,\tau)\int^\infty_0 dk k^2 e^{-(k^2/k_*^2(t))}
=(2\pi)^{-3/2}
 P_\chi(0,\tau) k_*^3(t)
, \dlabel{classvev} \ee
and using the convolution theorem we have \cite{p10} for $k\ll k_*$\footnote
{As usual $P_\chi\equiv (2\pi^2/k^3)\calp_\chi$, with both
$P$ and $\calp$ referred to as the spectrum.}
\be \calp_{\delta\chi^2}(t,k) = \frac1{\sqrt\pi} \vev{\chi^2(t)}^2 [k/k_*(\tau)]^3
, \dlabel{pchisq} \ee
with $\calp_{\delta\chi^2}(\tau,k)$ falling exponentially at $k\gg k_*$.

By virtue of \eq{adcon3}, the change in $\mmt$ in time $\mmt\mone$ is negligible
and so is the change in $a$. Setting $t=t_1+\mmt\mone$ we get
$k_*^2(t)\simeq a^2(t)\mmts$. But at this epoch the growth  era has hardly begun,
and subsequently $k_*(t)$ decreases while $a\mmt$ increases. So we really have
$k_*^2(t)\ll  a^2\mmts$ during the growth era, and $\chi_k$ falls exponentially
in the regime $k_*(t) < k < a\mmt$ which means that $k_*(t)$ is the dominant
mode.

We are mostly interested in an epoch just before the end of the linear era,
which we denote by a subscript $\svev$. We denote $k_*(t\svev)$ simply by $k_*$.
Let us define  $N\svev\equiv H(t\svev - t_1)$. 
If $N\svev\lsim 1$,
\be
1\lsim \frac{ \modm{t_1} }{ H} \lsim \( \frac{k_*}{a(t\svev)H} \)^2
\lsim \frac{ \modm{t\svev} }{ H} \leq  \frac mH\qquad (N\svev\lsim 1)
. \ee
If instead $N\svev\gsim 1$,  $k_*(t)$ levels out after $H(t-t_1)\sim 1$. We therefore
have, whatever the value of $N\svev$,
\be
e^{-N\svev} \lsim \( \frac {k_*}{aH} \)^2 \lsim \frac mH e^{-N\svev}
.\ee

Since $\mmt\gg H$, the upper bound on $k_*$ imples $(k_*/a) \chi_k \ll \dot\chi_k
=\mmt \chi_k$.
 The spatial gradient of $\chi(\bfx,t)$ will therefore be
small compared with the time-derivative, except near places where $\chi=0$.
Hence 
\bea \rho_\chi&\simeq& -\frac12 \mmts \chi^2 +\frac12 \dot \chi^2 \simeq 0 \\
 p_\chi &\simeq&  \frac12 \mmts \chi^2 +\frac12 \dot \chi^2
\simeq  \mmts \chi^2  \simeq \dot\chi^2
. \dlabel{pchi2} \eea
To evaluate $\rho_\chi$ we can use the local continuity equation
(valid because the spatial gradient is negligible), 
\be
\dot\rho_\chi= - 3H(\rho_\chi+ p_\chi) \simeq - 3H p_\chi
, \ee
giving
\be
\rho_\chi \simeq -(3H/2\mmt) p_\chi
. \ee

\section{Curvature perturbation}

\dlabel{scurv}

Now we calculate the contribution of $\chi$ to the curvature perturbation,
 following closely the procedure of \cite{p10}.
The curvature perturbation is given by $\zeta(\bfx,t)=\delta N(\bfx,t)$,
where $N$ is the number of $e$-folds of expansion from any initial
 slice with  $a(\bfx,t)=a(t)$ (flat slice) to a slice of uniform $\rho$
at time $t$. The contribution of $\chi$ during the linear era is $\zeta_\chi(\bfx)
=\delta N(\bfx,t\svev,t_1)$, where $N(\bfx,t,t_1)$ is the expansion from a slice of
uniform $\rho$ just after the beginning of the linear era, 
 to a slice of uniform $\rho$ just before its end. 
We   are working
in a gauge where $\delta\phi=0$ so that $\rho(\bfx,t)=\rho_\chi(\bfx,t) + \rho_\phi(t)$.
Since  $|\delta\rho_\chi(\bfx,t_1)|\ll |\delta\rho_\chi(\bfx,t\svev)|$ 
we have
\be
\zeta_\chi(\bfx,t) = - H  \frac{ 
\delta\rho_\chi(\bfx,t\svev) 
}{ 
\dot\rho(t)}
= \frac13 \frac{ \delta\rho_\chi(\bfx,t\svev) 
}{ 
\vev{\dot\chi^2(t\svev)} +\dot\phi^2(t) }
, \ee
where $\rho(t)$ is the spatial average of $\rho(\bfx,t)$.

Using the equations of the previous section this gives for $k\ll k_*$
\be
\zeta_\chi(\bfx) = -\frac {H}{2\modm{t\svev} }
\frac{
\vev{\dot\chi^2(t\svev)}
}{
\vev{\dot\chi^2(t\svev)} + \dot\phi^2(t\svev) } 
 \frac{
\delta \chi^2(\bfx,t\svev)
}{
\vev{\chi^2(t\svev)} 
} \dlabel{zetachi}
, \ee 
and
\be
\calp_{\zeta_\chi}(k) \simeq  \[ \frac H{2\modm{t\svev} }
\frac{
\vev{\dot\chi^2(t\svev) }
}{
\vev{\dot\chi^2(t\svev)} + \dot\phi^2(t\svev)
} \]^2
\( \frac k{k_*} \)^3
. \dlabel{pzetachi} \ee
At $k\gg k_*$, $\calp_{\zeta_\chi}$ is negligible  because $\chi_k$ is.
The spectrum therefore peaks at $k\sim k_*$.

If   $\vev{\dot\chi^2(t\svev)} \gg \dot\phi^2(t\svev)$
we have $\dot\rho\simeq \dot\rho_\chi$, which means that the slice of uniform
$\rho$ is practically the same as the one of uniform $\rho_\chi$. In turn, that
is the same as the slice of uniform $\chi$. 

In general the formula  $\zeta=\delta N$  holds only
 after smoothing $\rho$ and $p$ on a scale big enough that the local
continuity equation is satisfied.  In our case though, that equation is satisfied on all
of the  scales $k\gg k_*$ on which $\zeta_\chi$ is significant. Therefore,
the formula $\zeta_\chi(\bfx)=\delta N(\bfx,t\svev,t_1)$  makes sense on all of these scales. 
In the opposite 
regime  $k\ll k_*$, $\zeta_\chi$ will be negligible simply because $\delta\chi^2$
is, its  exponential growth not having begun. Thefore, $\calpzchi$ peaks at $k_*$
whether that scale is super-horizon or sub-horizon.

We want to see whether  the black hole bound of Section \ref{sbh} is satisfied
by \eq{pzetachi}. The situation is simple if  $k_*$ is super-horizon. Then,  
 $\zeta$ is of the form \eq{zetang} with the minus sign, and the
 black hole bound is $\calpzchi(k_*)< 6$ which is well satisfied.

If instead $k_*$ is sub-horizon, we have to remember that the black hole bound
refers to  horizon-sized regions. To apply it, we must 
 drop sub-horizon  modes of $\zeta_\chi$. Estimating the bispectrum, trispectrum 
as in \cite{p10}, one sees that this makes $\zeta_\chi$ nearly gaussian.
The black hole bound is  therefore roughly
$\calpzchi(H)\lsim 10\mtwo$. Since $k_*\gg k\sub{end}$ and $\mmt\gg H$,
it too is presumably  satisfied.

Earlier calculations of $\zeta_\chi$ are reviewed in \cite{p10}, including
the one in \cite{gs}. A continuation of \cite{gs} has since appeared \cite{afs}.
These papers consider the potential $V(\phi)\propto \phi^2$, for which 
$m^2(t) \propto t $ is a good approximation. Their expression for $\zeta_\chi$
(Eq.\ A(10) of
\cite{gs} and Eq.\ (6.15) of \cite{afs}) is the same as our \eq{zetachi},
except that the middle factor is missing. However, the meaning of
$\delta\chi$ and $\vev{\chi^2}$ in the final factor is different in their expression
from that in \eq{zetachi}; $\chi$ in their case is smoothed on the horizon scale
whereas in our case it is not. 

\section{Conclusion}

\dlabel{sconc}

We have calculated the contribution to $\zeta$ generated during the 
linear era of the waterfall, assuming that such an era exists and that
exponential growth takes place during that era. Such growth can occur only
if the tachyonic mass of the waterfall field is $m\gg H$ (standard hybrid inflation)
which we therefore demand.

The calculation generalises \cite{p10}, to the case of an arbitrary time-dependence
of $m^2(t)$ subject to it being slow enough to allow exponential growth.
Because the dependence is arbitrary, we cannot repeat the detailed investigation
given in \cite{p10}, of the parameter space within which our calculation will
apply. That could only be done with a specific inflaton potential $V(\phi)$, 
which among other things would determine $m^2(t)$.

We have, as usual,   considered only super-horizon modes of the
 curvature perturbation $\zeta$. The usual 
super-horizon treatment of $\zeta$, based on the local energy continuity
equation does still hold for   sub-horizon modes  during the waterfall.
But we have not used these modes, and to do so would require evolution
of the cosmological perturbations through to the post-inflation era.
In particular, that would be necessary if we were to discuss the possible formation of
black holes whose size is much smaller than the horizon scale at the end of 
inflation.

%%%%%%%%%%%%%%%%%%%%%%%%%%%%%%%%%%%%%%%%%%%%%%%%%%%%%%%%%%%%%%%%%%%%%%
\section{Acknowledgments}
%%%%%%%%%%%%%%%%%%%%%%%%%%%%%%%%%%%%%%%%%%%%%%%%%%%%%%%%%%%%%%%%%%%%%%
The author acknowledges support from
 UNILHC23792, European Research and Training Network (RTN) grant.

%%%%%%%%%%%%%%%%%%%%%%%%%%%%%%%%%%%%%%%%%%%%%%%%%%%%%%%%%%%%%%%%%%%%%%

\end{document}